\documentclass[aps,prl,nofootinbib,twocolumn,floatfix,superscriptaddress]{revtex4}
\def\mysection#1{{\bf #1.} }

\usepackage{epsfig}

\newcommand{\be}{\begin{equation}}
\newcommand{\ee}{\end{equation}}
\newcommand{\bea}{\begin{eqnarray}}
\newcommand{\eea}{\end{eqnarray}}
\newcommand{\beq}{\begin{equation}}
\newcommand{\eeq}{\end{equation}}
\def\beqa{\begin{eqnarray}}
\def\eeqa{\end{eqnarray}}
\newcommand{\no}{\nonumber}
\def\lsim{\mathrel{\rlap{\lower4pt\hbox{\hskip1pt$\sim$}}
    \raise1pt\hbox{$<$}}}         %less than or approx. symbol
\def\gsim{\mathrel{\rlap{\lower4pt\hbox{\hskip1pt$\sim$}}
    \raise1pt\hbox{$>$}}}         %greater than or approx. symbol

\newcommand{\slep}{{\tilde{l}}}
\newcommand{\ES}{E_{{\rm shift}}}

\newcommand{\gev}{{\rm GeV}}
\newcommand{\pb}{{\rm pb}}

\newcommand{\ifb}{{\rm fb}^{-1}}

\begin{document}

%\preprint

\preprint{UCI-TR-2009-10}
\preprint{CAVENDISH-HEP-09/10}

\title{\boldmath The Shifted Peak: Resolving Nearly Degenerate
  Particles at the LHC}

\author{Jonathan L.~Feng}
\affiliation{Department of Physics and Astronomy, University of
California, Irvine, CA 92697, USA }

\author{Sky T.~French}
\affiliation{Cavendish Laboratory, J.~J.~Thomson Avenue, Cambridge,
  CB3 0HE, UK }

\author{Christopher G.~Lester}
\affiliation{Cavendish Laboratory, J.~J.~Thomson Avenue, Cambridge,
  CB3 0HE, UK }

\author{Yosef Nir}
\affiliation{Department of Particle Physics,
  Weizmann Institute of Science, Rehovot 76100, Israel}

\author{Yael Shadmi}
\affiliation{Physics Department, Technion--Israel Institute of
Technology, Haifa 32000, Israel }

\vspace*{1cm}
%\date{\today}
%\pacs{12.10.Dm, 12.10.Kt, 98.80.Cq}

\begin{abstract}
We propose a method for determining the mass difference between two
particles, $\slep_1$ and $\slep_2$, that are nearly degenerate, with
$\Delta{m}\equiv m_2-m_1\ll m_1$.  This method applies when (a) the
$\slep_1$ momentum can be measured, (b) $\slep_2$ can only decay to
$\slep_1$, and (c) $\slep_1$ and $\slep_2$ can be produced in the
decays of a common mother particle.  For small $\Delta{m}$, $\slep_2$
cannot be reconstructed directly, because its decay products are too
soft to be detected.  Despite this, we show that the existence of
$\slep_2$ can be established by observing the shift in the mother
particle invariant-mass peak, when reconstructed from decays to
$\slep_2$.  We show that measuring this shift would allow us to
extract $\Delta{m}$.  As an example, we study supersymmetric
gauge-gravity hybrid models in which $\slep_1$ is a meta-stable charged
slepton next-to-lightest supersymmetric particle and
$\slep_2$ is the next-to-lightest slepton with
$\Delta{m} \sim 5~\gev$.
\end{abstract}

\maketitle

%%%%%%%%%%%%%%%%%%%%
\mysection{Introduction}
How well can we measure the masses of new particles at the LHC?  This
question might seem secondary at the early stages of discovery.  It
becomes significant, however, if there are two new particles that are
nearly degenerate, with a mass difference of the order of a few GeV.
In the busy LHC environment, it will be highly non-trivial to observe
such mass differences. On the other hand, it is quite natural to
expect such scenarios. If the new physics replicates the
three-generation structure of the standard model, then we could easily
have particles with the same gauge charges whose mass differences only
depend on small ``flavor effects.''  Obvious examples are the squarks
and sleptons of supersymmetry, particularly the first- and
second-generation ones.  Here we propose an indirect method for
resolving such mass differences, which applies when (a) the momentum
of the lighter particle can be measured, (b) the heavier particle can
only decay to the lighter particle, and (c) the two particles can be
produced in the decays of a common mother particle.  For concreteness,
we will discuss the case of two sleptons, $\slep_{1,2}$, with
$\Delta{m}\equiv m_2-m_1 \ll m_1$, with a meta-stable $\slep_1$.  Such
a spectrum is predicted by supersymmetric
models~\cite{Feng:2007ke,Nomura:2007ap,Hiller:2008sv} that explain the
masses and mixings of the standard model charged leptons and neutrinos
in terms of broken flavor symmetries~\cite{Froggatt:1978nt}. More
generally, this scenario is often realized in gauge-mediated
supersymmetry breaking models~\cite{Feng:1997zr}, and in large regions
of the parameter space of models with gravity-mediated supersymmetry
breaking~\cite{Feng:2003xh}.  Since the meta-stable $\slep_1$ leaves a
track in the muon detector, its momentum can be measured, and the
event is fully reconstructible.  Similar spectra and phenomena are
also possible in other frameworks, such as for the Kaluza-Klein
excitations of quarks and leptons in models with universal extra
dimensions~\cite{Appelquist:2000nn}.

The method relies on the decays of a mother particle, in this case the
neutralino $\chi^0_1$, to $\slep_{1,2}$.  If we use direct decays to
$\slep_1$ to reconstruct the neutralino, the invariant mass
distribution will be peaked at the correct neutralino mass.  Some of
the time, however, the neutralino decays to $\slep_2$, which
subsequently decays to $\slep_1$. The leptons produced in the $\slep_2
\to \slep_1$ decay are relatively soft, with energies typically of the
order of $\Delta{m}$.  They may therefore be lost, implying that the
decays $\chi^0_1 \to \slep_1$ and $\chi^0_1 \to \slep_2$ have the same
topology.  Rather than blurring the picture, however, we show that by
attempting to reconstruct the $\chi^0_1$ in both cases, one can in
fact find two peaks: one at the neutralino mass $M$, and one slightly
shifted to a lower value by an amount $\ES \sim \Delta{m}$.  Thus,
measuring a shift in the neutralino mass peak will tell us that there
are in fact two slepton states lighter than the neutralino, with a
mass difference roughly given by $\ES$.  Furthermore, if the two
particles in question are scalars (which can be established from the
angular distribution of the decays), $\Delta{m}$ can be determined in
terms of $\ES$, $m_1$, and the neutralino mass $M$.

%%%%%%%%%%%%%%%%%%%%
\mysection{The Shifted Peak}
By assumption, the neutralino has two possible decays into sleptons.
The first is the direct decay to $\slep_1$,
\begin{equation}\label{eq:ee}
\chi_1^0\to\slep_1^\pm l_1^\mp \ .
\end{equation}
The second is the decay to $\slep_2$,
\begin{equation}\label{eq:mm}
\chi_1^0\to \slep_2^\pm l_2^\mp \ ,
\end{equation}
followed by one of the two three-body
decays~\cite{Ambrosanio:1997bq,Feng:2009bs}
\begin{eqnarray}\label{eq:xz}
\slep_2^\pm&\to& \slep_1^\pm X^{\pm \mp} \ , \\
\slep_2^\pm&\to& \slep_1^\mp X^{\pm\pm}\ , \label{eq:xc}
\end{eqnarray}
where $X^{\pm \mp}$ contains two opposite-sign (OS) leptons, and
$X^{\pm\pm}$ contains two same-sign (SS) leptons.  Note that the
charge-flipping decays of Eq.~(\ref{eq:xc}) resulting in SS leptons
are possible because the neutralino is a Majorana fermion; SS leptons
are also present in models other than supersymmetry when the decay is
mediated by a vector boson or a scalar. We denote these lepton pairs
by $X$ to emphasize the fact that they are too soft to pass our cuts.
Thus, the observed particles are the hard lepton from
Eq.~(\ref{eq:ee}) or (\ref{eq:mm}), and the long-lived slepton
$\slep_1$ from Eq.~(\ref{eq:ee}), (\ref{eq:xz}), or (\ref{eq:xc}).

We can thus construct distributions for the following invariant
masses-squared:
\begin{eqnarray}
m_{\slep l_1}^2&\equiv& \left(p_{\slep_1} +p_{l_1}\right)^2 \ ,\no\\
m_{\slep l_2}^2&\equiv& \left(p_{\slep_1} +p_{l_2}\right)^2 \ ,
\end{eqnarray}
where the $\slep_1$ and $l_2$ charges can be either opposite or the
same.  Obviously,
\begin{equation}\label{eq:mee}
m_{\slep l_1}=M \ ,
\end{equation}
where $M$ is the neutralino mass.  However, because of the missing
soft leptons,
\begin{equation}
m_{\slep l_2}\neq M \ ,
\end{equation}
and the peak of the $m_{\slep l_2}$ distribution is shifted from $M$
to $M-\ES$.

The identities of $l_1$ and $l_2$ depend, of course, on the flavor
decompositions of $\slep_1$ and $\slep_2$, respectively.  In one
extreme case, if $\slep_{1,2}$ are the left- and right-handed sleptons
associated with the same flavor, the two leptons are identical.
Still, a peak in the invariant mass $m^2_{\slep_1^\pm l_2^\pm}$,
formed from events with SS sleptons and leptons, can only come from
the decays of Eq.~(\ref{eq:mm}) and will therefore exhibit the shift
$\ES$. The analogous OS distribution will contain both types of events
specified in Eqs.~(\ref{eq:ee}) and~(\ref{eq:mm}), and so will exhibit
a double peak structure, with the two peaks separated by $\ES$.

In the opposite extreme, the two sleptons could be pure states of
different flavors.  In this case, the leptons $l_1$ and $l_2$ are
different flavors, and there is no need to rely on the charges to
separate the distributions.  In any case, we can use the shifted peak
to infer the existence of two distinct states. In the following, for
the sake of simplicity, we will consider the second case, and take
$\slep_1=\tilde{e}$ and $\slep_2=\tilde{\mu}$ so that $l_1=e$ and
$l_2=\mu$.

We now turn to the calculation of the peak shift $\ES$.  We denote the
four momentum of a particle $a$ by $p_a$, and the $\slep_1$ energy by
$E_1$.  Then,
\begin{equation}\label{eq:memu}
m^2_{{\slep\mu}}=M^2-m_2^2 +m_1^2 -2p_\mu\cdot p_X \ .
\end{equation}
{}From here on, we neglect the lepton masses.  Working in the $\slep_2$
rest frame, we take the $x-y$ plane to be the plane of the muon and
dilepton momenta, $\vec p_\mu$ and $\vec p_X$.  We further take the
muon direction to define the $\hat x$-axis.  The four-momenta of the
hard muon and of the soft dilepton are then given by
\begin{eqnarray}\label{eq:pmu}
p_\mu&=&\frac{M^2-m_2^2}{2m_2}(1,\hat x)\ , \\
\label{eq:pX}
p_X&=&\left( m_2-E_1,-\hat n\sqrt{{E_1}^2-m_1^2} \right) \ ,
\end{eqnarray}
where $\hat n=(\cos\theta, \sin\theta, 0)$.  Substituting
Eqs.~(\ref{eq:pmu}) and (\ref{eq:pX}) into Eq.~(\ref{eq:memu}), we find
\begin{equation}
m^2_{\slep\mu}=m_1^2+\frac{M^2-m_2^2}{m_2}\left(E_1
-\cos\theta\sqrt{E_1^2-m_1^2}\right) ,
\end{equation}
with $E_1$ and $\cos\theta$ varying {\em independently} in the
intervals
\begin{eqnarray}\label{emax}
E_1 &\in& \left[ m_1, \frac{m_2^2+m_1^2}{2m_2}\right] ,\\
\cos\theta &\in& [-1, +1] \ .
\end{eqnarray}

Let us now consider these quantities for $\Delta{m}\ll m_1$.  Working
to leading order in $\Delta{m}$, we find that the maximum value of
$E_1$, given in Eq.~(\ref{emax}), is
\begin{equation}
\frac{m_2^2+m_1^2}{2m_2} \approx m_1\,\left(1+\frac12x^2\right) ,
\end{equation}
where
\begin{equation}
x \equiv \frac{\Delta{m}}{m_1}\ .
\end{equation}
We can therefore parameterize
\begin{equation}
E_1=m_1 \left(1+ \frac12 a x^2\right) ,
\end{equation}
where $0\leq a\leq1$ varies from event to event.  Note that the ${\cal
O}(x)$ piece vanishes.  To leading order in the mass splitting, we then
find
\begin{equation}
\label{eq:mlmudistribution}
m^2_{\slep\mu} -M^2 \sim
- \left[
(M^2 + m_1^2) + \cos\theta \sqrt{a} (M^2-m_1^2) \right] x\ .
\end{equation}
In reality, $\ES$ is defined by the peak of the $m^2_{\slep\mu}$
distribution to be
\begin{equation}
\ES=M-\sqrt{m^2_{\slep\mu}\vert_{{\rm peak}}} \ ,
\end{equation}
and so it depends on the matrix elements governing the decays.  Still,
recalling that $a \leq 1$, the second term on the right-hand side of
Eq.~(\ref{eq:mlmudistribution}) is always smaller than the first term
by at least
\begin{equation}\label{ratio}
\frac{M^2-m_1^2}{M^2+m_1^2} \ .
\end{equation}
Thus, a rough estimate for the mass splitting $\Delta{m}$
can be obtained from
\begin{equation}\label{shift}
\ES \sim \frac{M^2+m_1^2}{2Mm_1}\, \Delta{m} \ .
\end{equation}
Since the ratio of Eq.~(\ref{ratio}) will be measured, we will know
the accuracy of this estimate.

%%%%%%%%%%%%%%%%%%%%%%%%%%%%%%%%%%%
\mysection{Beyond Leading Order}
In fact, for a scalar $\slep_2$, we can carry out the analysis exactly
(including terms higher order in $\Delta m$), and for all possible
matrix elements, since we can determine the $m_{\slep\mu}^2$ peak
position based solely on kinematics.

The crucial point is that, since the sleptons are scalars, the
$\slep_2\to\slep_1 X$ decays are uniformly distributed in
$\cos\theta$. For a fixed $E_1$, then, these distributions are
centered at
\begin{equation}
m_{\slep\mu}^2 = m_1^2 + \frac{M^2 - m_2^2}{m_2} E_1 \ ,
\end{equation}
with width
\begin{equation}
\Delta(m_{\slep\mu}^2)=2\frac{M^2-m_2^2}{m_2}\sqrt{E_1^2-m_1^2} \ .
\end{equation}
As $E_1$ decreases from $(m_2^2+m_1^2)/(2m_2)$ to $m_1$, the width
gets smaller and smaller, until the distribution becomes infinitely
thin and centered at
\begin{equation}
\widehat{m}_{\slep\mu}^2=m_1^2+\frac{M^2-m_2^2}{m_2}m_1 =
M^2-\frac{M^2+m_1 m_2}{m_2} \,\Delta{m}\ .
\end{equation}
The total distribution is obtained by adding up all the contributions
from different values of $E_1$ with the appropriate weights.

We reach the following two conclusions. First, the peak of the total
distribution is at $\widehat{m}_{\slep\mu}^2$.  To see this, note that
\begin{eqnarray}
\widehat{m}_{\slep\mu}^2 \in
\left[ m_1^2 + \frac{M^2 - m_2^2}{m_2} \left(
E_1 - \sqrt{E_1^2 - m_1^2} \right),\right.&&\no\\
\ \ \ \left.m_1^2 + \frac{M^2 - m_2^2}{m_2} \left(
E_1 + \sqrt{E_1^2 - m_1^2}\right)\right]&&
\end{eqnarray}
for all possible $E_1$. Thus every $E_1$ contributes to this value of
$m_{\slep\mu}^2$, and this is the only value of $m_{\slep\mu}^2$ that
every $E_1$ contributes to.  It is therefore the peak.

Second, since the distribution is not symmetric about the peak, the
mean of the distribution need not be at $\widehat{m}_{\slep\mu}^2$.
This is relevant, because it means that the peak after experimental
smearing need not be at $\widehat{m}_{\slep\mu}^2$. However, the mean
must satisfy
\begin{equation}
\overline{m}_{\slep\mu}^2\in\left[ \widehat{m}_{\slep\mu}^2 ,
m_1^2 + \frac{(M^2 - m_2^2)(m_2^2 + m_1^2)}{2 m_2^2}  \right].
\end{equation}
Thus we see that, to first order in $\Delta{m}/m_2$, the peak and the
mean coincide, reproducing the result of Eq.~(\ref{shift}).

%%%%%%%%%%%%%%%%%%%%
\mysection{The Analysis}
We now illustrate our method by simulating events in a concrete
example model.  Rather than simulate a realistic supersymmetric model,
we use \textsc{Herwig}~\cite{Marchesini:1991ch,Corcella:2000bw} to
specify a model to isolate the processes we are interested in,
namely superpartners produced through $\tilde{q}
\tilde{q}$, $\tilde{q} \tilde{g}$, and $\tilde{g} \tilde{g}$ pair
production, followed by the cascade decays $(\tilde{g} \to) \tilde{q}
\to \chi^0_1 \to \tilde{l}_{1,2}$. To do this, we
choose $m_{\tilde{g}} = 650~\gev$, all squarks degenerate with
$m_{\tilde{q}} = 450~\gev$,
\begin{eqnarray}
M = m_{\chi_1^0}  &=& 225.2~\gev \ , \no\\
m_2 = m_{\tilde{l}_2} &=& 139.9~\gev \ , \\
m_1 = m_{\tilde{l}_1} &=& 134.9~\gev \ ,\no
\end{eqnarray}
and all other superpartners very heavy, so that they are decoupled
from collider events.  As mentioned above, it suffices for our
purposes to consider flavor-diagonal soft terms and neglect left-right
slepton mixing.  We then let $\tilde{l}_{1} = \tilde{e}_R$ and
$\tilde{l}_{2} = \tilde{\mu}_R$, and set $B(\chi^0_1 \to \tilde{l}_1)
= B(\chi^0_1 \to \tilde{l}_2)$, as appropriate for the case of a
gaugino $\chi_1^0$ and right-handed sleptons.

We generate events for a $\sqrt{s} = 14~\gev$ LHC and pass these
events through a generic LHC detector simulation, \textsc{AcerDET
1.0}~\cite{RichterWas:2002ch}. We configure \textsc{AcerDET} as
follows: electrons and muons are selected with $p_T>6~\gev$ and
$|\eta|<2.5$.  Electrons and muons are considered to be isolated if
they lie at a distance greater than $\Delta R>0.4$ from other leptons
or jets and if less than 10 GeV of energy is deposited in a cone of
$\Delta R = 0.2$, where $\Delta R = \sqrt{(\Delta\eta)^2 +
(\Delta\phi)^2}$.  The lepton momentum resolutions we use are
parameterized from the results of Full Simulation of the ATLAS
detector~\cite{Aad:2008zzm}. (Our electrons are smeared according to a
pseudo-rapidity-dependent parameterization, whilst muons are smeared
according to the results for $|\eta|<1.1$). \textsc{AcerDET} does not
take into account lepton reconstruction efficiencies. We therefore
apply by hand a reconstruction efficiency of $90\%$ to the muons and a
reconstruction efficiency of $77\%$ to the electrons. This gives
$0.86$ as the ratio of electron to muon reconstruction efficiency. We
generate 20,000 events (before any cuts or requirements are imposed),
which corresponds to 40,000 $\chi^0_1$ decays. The supersymmetry cross
section for the events of interest and for the parameters we have
chosen is $\sim 50~\pb$, and so our event samples assume an integrated
luminosity of ${\cal L} \sim 0.4~\ifb$.  In a more realistic model, only
a fraction $\epsilon$ of all supersymmetry events will satisfy our
event criteria, and so the assumed luminosity is $0.4~\ifb/\epsilon$.

The ATLAS trigger and reconstruction have been modified recently to
include the possibility of triggering on and reconstructing
meta-stable sleptons~\cite{Tarem:2009zz}.
We conservatively restrict ourselves to consideration of sleptons
traveling fast enough to arrive in the principal time bin (25~ns
wide).  Furthermore, as mass resolution degrades as $\beta$ approaches
unity, it is necessary to place an upper limit on $\beta$.  Combining
these two requirements we demand that the $\slep_1$ candidates have
velocity $0.6<\beta<0.8$ in agreement with Ref.~\cite{Ellis:2006vu}.
Note that the upper cut on $\beta$ will also greatly reduce
backgrounds~\cite{CMSNote}; we assume here that the remaining
background events do not impact our results significantly.

For the $\slep_1$, we take their momenta
$|\vec p_{\tilde l_1}|$ from truth and  smear them by a Gaussian
with $\sigma=0.05|\vec p_{\tilde l_1}|$. We also smear the
$\beta$ by a Gaussian with $\sigma=0.02$. These choices are again
motivated by the results of Ref.~\cite{Ellis:2006vu} and their
reconstruction of the $\tilde l_1$ by measuring slepton time of flight
with the ATLAS RPC chambers. We assume that $m_1$ is well measured, so
we scale the four-momentum components to give the exact value of the
mass. The resulting $\tilde{l}_{1}$ mass distribution is given in
Fig.~\ref{fig:SingleSlepton}.  We now take our measurement of $m_1$ to
be 134.9 GeV. (Although our Gaussian fit actually gives $135.0\pm0.1$
GeV, we assume that there is no systematic which would prevent us from
obtaining $134.9$ GeV exactly.)
\begin{figure}
\centerline{\psfig{file=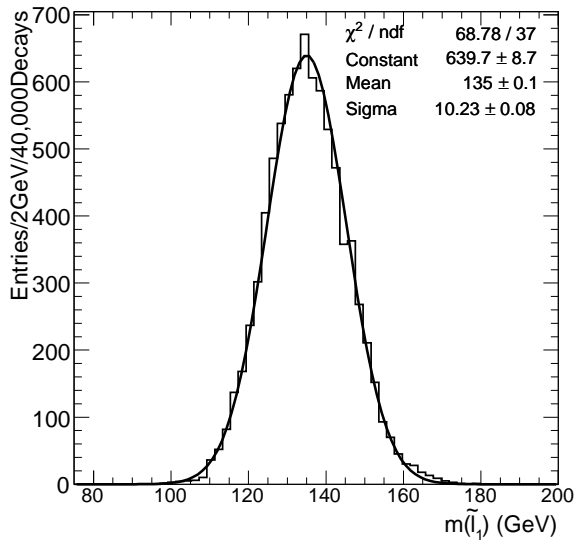,width=1\linewidth}}
\vspace*{-.3in}
\caption{\label{fig:SingleSlepton} Reconstructed $\tilde l_{1}$ with
$0.6<\beta<0.8$ after smearing $\tilde l_{1}$ momenta with a Gaussian
with $\sigma = 0.05|\vec p_{\tilde l_1}|$ centered on the true momentum 
and after smearing $\beta$ with a Gaussian with $\sigma = 0.02$.  }
\end{figure}

To eliminate the soft leptons from $\tilde{l}_2$ decays, we impose a
hard $p_T$ cut of 30 GeV. The $p_T$ distribution of these soft leptons
is shown in Fig.~\ref{fig:SoftLeptons}, from which we learn that such
a cut is indeed a reasonable choice. Of course, in reality, the
information presented in this Figure will not be available, and the
experimenters will have to work by trial and error to find an optimal
cut that gives the largest number of peak events while maintaining a
clean peak.

We first consider all OS $\slep_1^\pm e^\mp$ and
$\slep^\pm_1 \mu^\mp$ events and reconstruct the invariant masses
$m_{\slep_1 e}$ and $m_{\slep_1\mu}$, imposing the 30 GeV $p_T$ cut on
leptons.
Based on Eq.~(\ref{shift}), we expect $\widehat m_{\slep_1
e}-\widehat{m}_{\slep_1\mu}=5.6$~GeV:
\begin{eqnarray}
\widehat m_{\slep_1 e}&=&225.2~\gev \ ,\no\\
\widehat{m}_{\slep_1 \mu}&=&219.6~\gev \ .
\end{eqnarray}
The invariant mass distributions are shown in
Figs.~\ref{fig:OS_eslep} and \ref{fig:OS_muslep}.  We decompose each
of these distributions into two pieces by fitting them to the sum of
an exponentially falling contribution and a Gaussian distribution,
with form
\begin{equation}
\label{eq:fit}
\frac{dN}{dm} = N_{\rm tot}
\left[ (1- f_{\rm sig}) |a| e^{a m}
+ f_{\rm sig} \sqrt{\frac{2}{\pi}} \frac{1}{\sigma}
e^{-\frac{(m - {\rm mean})^2 }{2\sigma^2}} \right] \ ,
\end{equation}
where $a$ and ${\rm mean}$ have units of $\gev^{-1}$ and $\gev$,
 respectively.  These decompositions are also shown in
 Figs.~\ref{fig:OS_eslep} and \ref{fig:OS_muslep}.  The peaks of the
 Gaussian components then give us
\begin{eqnarray}\label{eq:data}
\widehat m_{\slep e}&=&225.4\pm0.1~\gev \ , \no\\
\widehat{m}_{\slep \mu}&=&219.2\pm0.3~\gev \ .
\end{eqnarray}
Thus, the separation between the peaks is experimentally well
established. Using Eq.~(\ref{shift}), we infer from the results of
Eq.~(\ref{eq:data}) that the selectron and the smuon are split in
mass, with
\begin{equation}
\Delta m=5.5\pm0.3 ~\gev \ ,
\end{equation}
to be compared with our input value of 5.0 GeV.  
The exponentially-falling background in Figs.~\ref{fig:OS_eslep} and
\ref{fig:OS_muslep} is purely combinatoric. It consists of
combinations of a $\slep_1$ with a lepton on the ``other side'' of the
event.  In this toy MC we neglected other sources of SUSY background,
from events with more than one lepton on each side of the decay.  One
can check however, (see Ref.~\cite{longpaper}) that these hardly
affect our results, since the decay chain we considered here is the
dominant one.

\begin{figure}
\centerline{\psfig{file=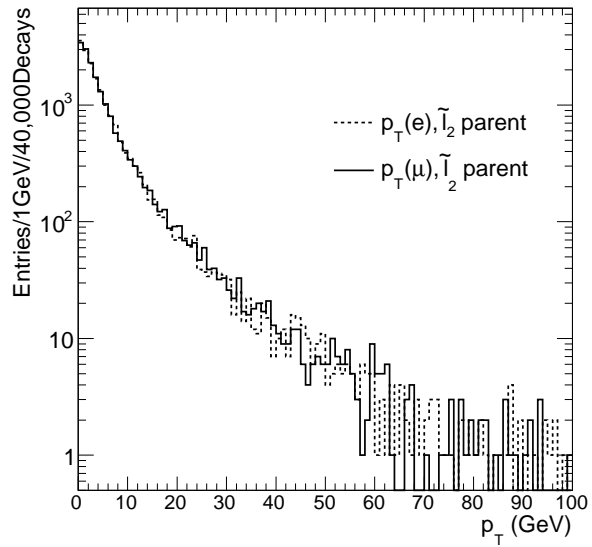,width=1\linewidth}}
\vspace*{-.3in}
\caption{\label{fig:SoftLeptons} The $p_T$ distribution of leptons
produced by the three-body decays of $\tilde l_2$.}
\end{figure}

\begin{figure}
\centerline{\psfig{file=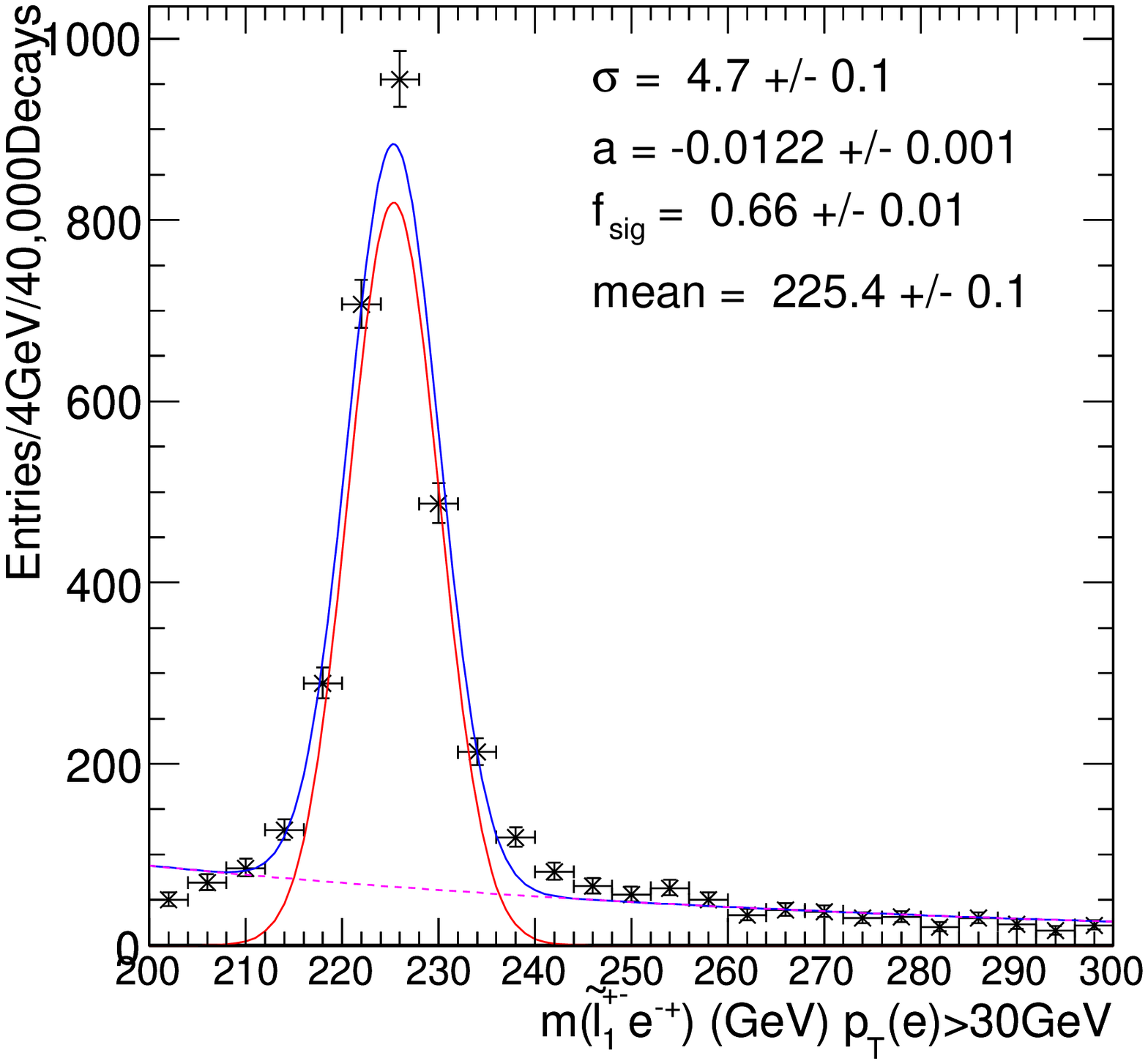,width=1\linewidth}}
\vspace*{-.3in}
\caption{\label{fig:OS_eslep} The $\tilde l^{\pm}_{1} e^{\mp}$
invariant mass distribution. The fit parameters $\sigma$, $a$,
$f_{\rm sig}$, and mean are defined in Eq.~(\ref{eq:fit}). }
\end{figure}
\begin{figure}
\centerline{\psfig{file=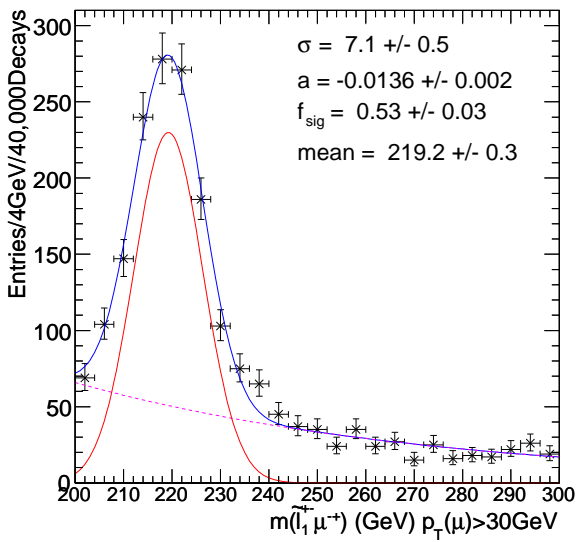,width=1\linewidth}}
\vspace*{-.3in}
\caption{\label{fig:OS_muslep} The $\tilde l^{\pm}_{1}\mu^{\mp}$
invariant mass distribution. The fit parameters $\sigma$, $a$,
$f_{\rm sig}$, and mean are defined in Eq.~(\ref{eq:fit}). }
\end{figure}

So far we considered events with OS $\slep_1$ and lepton.
It is also instructive to examine the equivalent SS invariant mass
distributions,  shown in Figs.~\ref{fig:SS_eslep} and \ref{fig:SS_muslep}.
As expected, we see no peak in the $\tilde l^\pm_1 e^\pm$ distribution,
and only the shifted peak in the $\tilde l_1^\pm \mu^\pm$ distribution.
Additionally we see that the muon peaks in the OS and SS
samples are of similar size.  This implies that the probability for
charge-preserving and charge-flipping decays are similar.  This is not
a given~\cite{Ambrosanio:1997bq,Feng:2009bs}, and so provides an
interesting additional constraint on supersymmetric models.
Note that, if the light sleptons are mixed flavor states, the SS distribution
can serve to reduce some of the background to the shifted peak.
With mixed states, Fig.~\ref{fig:OS_muslep} would contain contamination
from direct neutralino decays to $\slep_1$ and a muon, but these
must have opposite charges, and so do not contribute to the SS distributions.

\begin{figure}
\centerline{\psfig{file=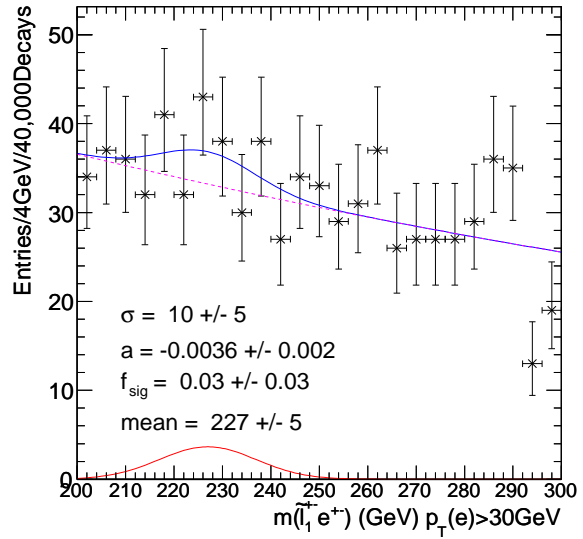,width=1\linewidth}}
\vspace*{-.3in}
\caption{\label{fig:SS_eslep} The $\tilde l^{\pm}_{1} e^{\pm}$
invariant mass distribution. The fit parameters $\sigma$, $a$,
$f_{\rm sig}$, and mean are defined in Eq.~(\ref{eq:fit}). }
\end{figure}
\begin{figure}
\centerline{\psfig{file=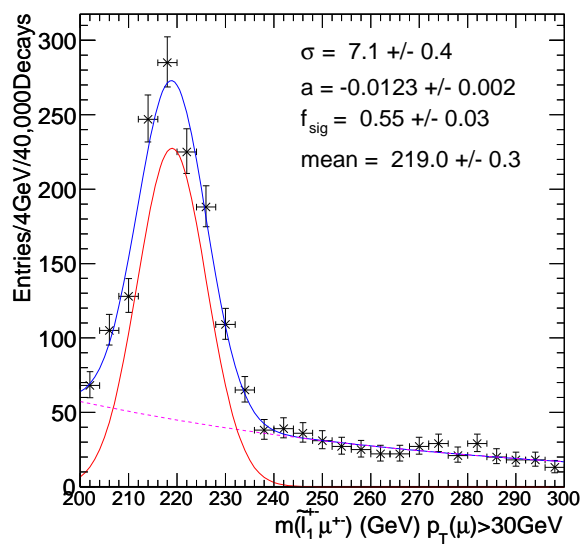,width=1\linewidth}}
\vspace*{-.3in}
\caption{\label{fig:SS_muslep} The $\tilde l^{\pm}_{1} \mu^{\pm}$
invariant mass distribution. The fit parameters $\sigma$, $a$,
$f_{\rm sig}$, and mean are defined in Eq.~(\ref{eq:fit}). }
\end{figure}

It would be nice to close this section by
reporting a minimum $\Delta m$ that, given our assumptions about the
experiment, could be established and measured.  The power to
discriminate peaks will depend on many things, however.  These
include, but are not limited to, the cross section in each peak, the
degree to which the peak shapes are modified by variation in
acceptance across each plot, the cross section of the underlying
backgrounds, and the uncertainties in those backgrounds.  As mentioned
above, with the strict upper cut on $\beta$ that we used, we do not
expect significant SM background. There would be however SUSY
background as well as cavern back-splash background (not modeled in
this investigation).  For the numbers of events simulated in this
study, and with not all sources of backgrounds simulated, it is clear
that the errors on the means fitted to the invariant mass peaks, being
of order $0.3$ GeV, would suggest that peak discrimination at the 5
sigma level is unlikely to be below 1 or 2 GeV. Conversely, one would
hope that discrimination at the level of 5 to 10 GeV (the width of the
reconstructed invariant mass peaks) ought to be possible, and is
indeed demonstrated here.

%%%%%%%%%%%%%%%%%%%%
\mysection{Conclusions}
We described a method for determining whether a newly discovered
long-lived particle is accompanied by another particle of almost
degenerate mass, and for measuring the mass difference between the
two.

In the case of supersymmetry, which is the focus of our discussion,
such measurements are central to analyzing the flavor structure of the
theory, which will tell us about the origin of supersymmetry breaking.
If the mechanism that mediates supersymmetry breaking is minimally
flavor violating (MFV)~\cite{D'Ambrosio:2002ex,Cirigliano:2005ck},
such as with pure gauge-mediation, then the mass splitting between the
selectron and the smuon is expected to be determined by the
muon-Yukawa squared, $\Delta m/m\lsim y_\mu^2\sim10^{-6}\tan^2\beta$,
below the percent level. If we can experimentally establish a larger
mass splitting, then we would obtain an intriguing clue for
contributions that are not MFV~\cite{Feng:2007ke}.

If the dominant mechanism of supersymmetry breaking is
gauge-mediation, then the lightest charged slepton can be long-lived
and leave a charged track in the muon detector of ATLAS/CMS.  If the
mass splitting is large enough that the transverse momenta of all the
resultant leptonic decay products are frequently above the
experimental reconstruction minimum (likely to be in the range 5 to 10
GeV) then we can fully reconstruct the $\slep_2\to\slep_1$ decays and
measure the mass splitting directly. If, however, the mass splitting
is only a few GeV, the leptons produced in this decay are typically
too soft to be detected.  Despite this, as we have shown here, the
$\slep_1 l$ invariant mass distribution will exhibit a double peak
structure.  The separation between the two peaks can be experimentally
established and measured, and its value can be translated into a value
of the slepton mass splitting.

We note that a related study was presented in
Ref.~\cite{Allanach:2008ib} in the context of gravity-mediated
supersymmetry breaking, with a neutralino lightest supersymmetric
particle.  The events are characterized by missing energy, and the
measurement of the slepton mass splitting is based on the kinematic
edges of the $m^2_{l^+l^-}$ distribution from $\chi_2^0 \to
\slep_{1,2}^\pm \ell^\mp \to \chi_1^0 \ell^\mp \ell^\pm$.

With non-MFV supersymmetry breaking, we expect that the slepton mass
eigenstates are not identical to the lepton flavor eigenstates.  In
this case, the decays we have considered contain a great deal of
flavor mixing
information~\cite{ArkaniHamed:1996au,Engelhard:2009br,Feng:2009bs},
and our method can be used to measure not only the mass splitting, but
also the mixing~\cite{longpaper}.  The SS sample will be particularly
useful in this case, as it is only sensitive to neutralino decays to
$\slep_2$, and therefore only exhibits the ``shifted peak.''  Counting
the number of electrons and muons in this shifted peak gives a clean
measurement of the flavor decomposition of $\slep_2$.

%%%%%%%%%%%%
\mysection{Acknowledgments}
We thank Kfir Blum, Jie Chen, James Frost, Iftah Galon, Are Raklev,
David Sanford, Shlomit Tarem, and Felix Yu for technical assistance and helpful
conversations.  The work of JLF was supported in part by NSF grant
PHY--0653656.  The work of YN is supported by the Israel Science
Foundation (ISF) under grant No.~377/07, the German-Israeli foundation
for scientific research and development(GIF), and the Minerva
Foundation.  The work of YS was supported in part by the Israel
Science Foundation (ISF) under grant No.~1155/07. This research was
supported in part by the United States-Israel Binational Science
Foundation (BSF) under grant No.~2006071.  CGL and STF acknowledge
support from the United Kingdom's Science and Technology Facilities
Council (STFC), Peterhouse and the University of Cambridge.

%\clearpage
%\vspace*{-5mm}
%%%%%%%%%%%%%%%%%%%%%


\begin{thebibliography}{99}

\bibitem{Feng:2007ke}
  J.~L.~Feng, C.~G.~Lester, Y.~Nir and Y.~Shadmi,
  %``The Standard Model and Supersymmetric Flavor Puzzles at the
  %Large Hadron Collider,''
  Phys.\ Rev.\  D {\bf 77}, 076002 (2008)
  [arXiv:0712.0674 [hep-ph]].
  %%CITATION = PHRVA,D77,076002;%%

\bibitem{Nomura:2007ap}
  Y.~Nomura, M.~Papucci and D.~Stolarski,
  %``Flavorful Supersymmetry,''
  Phys.\ Rev.\  D {\bf 77}, 075006 (2008)
  [arXiv:0712.2074 [hep-ph]].
  %%CITATION = PHRVA,D77,075006;%%

\bibitem{Hiller:2008sv}
  G.~Hiller, Y.~Hochberg and Y.~Nir,
  %``Flavor Changing Processes in Supersymmetric Models with Hybrid Gauge- and
  %Gravity-Mediation,''
  JHEP {\bf 0903}, 115 (2009)
  [arXiv:0812.0511 [hep-ph]].
  %%CITATION = JHEPA,0903,115;%%}

\bibitem{Froggatt:1978nt}
  C.~D.~Froggatt and H.~B.~Nielsen,
  %``Hierarchy Of Quark Masses, Cabibbo Angles And CP Violation,''
  Nucl.\ Phys.\  B {\bf 147}, 277 (1979).
  %%CITATION = NUPHA,B147,277;%%

\bibitem{Feng:1997zr}
  J.~L.~Feng and T.~Moroi,
  %``Tevatron signatures of longlived charged sleptons in gauge mediated
  %supersymmetry breaking models,''
  Phys.\ Rev.\  D {\bf 58}, 035001 (1998)
  [arXiv:hep-ph/9712499].
  %%CITATION = PHRVA,D58,035001;%%

\bibitem{Feng:2003xh}
  J.~L.~Feng, A.~Rajaraman and F.~Takayama,
  %``Superweakly-interacting massive particles,''
  Phys.\ Rev.\ Lett.\  {\bf 91}, 011302 (2003)
  [arXiv:hep-ph/0302215];
  %%CITATION = PRLTA,91,011302;%%
%\bibitem{Feng:2003uy}
%  J.~L.~Feng, A.~Rajaraman and F.~Takayama,
  %``SuperWIMP Dark Matter Signals from the Early Universe,''
  Phys.\ Rev.\  D {\bf 68}, 063504 (2003)
  [arXiv:hep-ph/0306024].
  %%CITATION = PHRVA,D68,063504;%%

\bibitem{Appelquist:2000nn}
  T.~Appelquist, H.~C.~Cheng and B.~A.~Dobrescu,
  %``Bounds on universal extra dimensions,''
  Phys.\ Rev.\  D {\bf 64}, 035002 (2001)
  [arXiv:hep-ph/0012100].
  %%CITATION = PHRVA,D64,035002;%%

\bibitem{Ambrosanio:1997bq}
  S.~Ambrosanio, G.~D.~Kribs and S.~P.~Martin,
  %``Three-body decays of selectrons and smuons in low-energy
  %supersymmetry breaking models,''
  Nucl.\ Phys.\  B {\bf 516}, 55 (1998)
  [arXiv:hep-ph/9710217].
  %%CITATION = NUPHA,B516,55;%%

\bibitem{Feng:2009bs}
  J.~L.~Feng, I.~Galon, D.~Sanford, Y.~Shadmi and F.~Yu,
  %``Three-Body Decays of Sleptons with General Flavor Violation
  %and Left-Right Mixing,''
  arXiv:0904.1416 [hep-ph].
  %%CITATION = ARXIV:0904.1416;%%

\bibitem{Marchesini:1991ch}
  G.~Marchesini, B.~R.~Webber, G.~Abbiendi, I.~G.~Knowles,
  M.~H.~Seymour and L.~Stanco,
  %``HERWIG: A Monte Carlo event generator for simulating hadron emission
  %reactions with interfering gluons. Version 5.1 - April 1991,''
  Comput.\ Phys.\ Commun.\  {\bf 67}, 465 (1992).
  %%CITATION = CPHCB,67,465;%%

\bibitem{Corcella:2000bw}
  G.~Corcella {\it et al.},
  %``HERWIG 6.5: an event generator for Hadron Emission Reactions With
  %Interfering Gluons (including supersymmetric processes),''
  JHEP {\bf 0101}, 010 (2001)
  [arXiv:hep-ph/0011363].
  %%CITATION = JHEPA,0101,010;%%

\bibitem{RichterWas:2002ch}
  E.~Richter-Was,
  %``AcerDET: A particle level fast simulation and reconstruction
  %package for
  %phenomenological studies on high p(T) physics at LHC,''
  arXiv:hep-ph/0207355.
  %%CITATION = HEP-PH/0207355;%%

\bibitem{Aad:2008zzm}
  G.~Aad {\it et al.}  [ATLAS Collaboration],
  %``The ATLAS Experiment at the CERN Large Hadron Collider,''
  JINST {\bf 3}, S08003 (2008).
  %%CITATION = JINST,3,S08003;%%

%\cite{Tarem:2009zz}
\bibitem{Tarem:2009zz}
  S.~Tarem, S.~Bressler, H.~Nomoto and A.~Di Mattia,
  %``Trigger and reconstruction for heavy long-lived charged particles with the
  %ATLAS detector,''
 Eur.\ Phys.\ J C DOI 10.1140/epjc/s10052-009-1040-0.

\bibitem{Ellis:2006vu}
  J.~R.~Ellis, A.~R.~Raklev and O.~K.~Oye,
  %``Gravitino dark matter scenarios with massive metastable charged
  %sparticles at the LHC,''
  JHEP {\bf 0610}, 061 (2006)
  [arXiv:hep-ph/0607261].
  %%CITATION = JHEPA,0610,061;%%

\bibitem{CMSNote}
See, for example,
CMS Collaboration, Physics Analysis Summary, CMS-EXO-08-003.

\bibitem{longpaper}
J.~L.~Feng, S.~T.~French, I.~Galon, C.~G.~Lester, Y.~Nir, D.~Sanford,
Y.~Shadmi and F.~Yu,
work in progress.

\bibitem{D'Ambrosio:2002ex}
  G.~D'Ambrosio, G.~F.~Giudice, G.~Isidori and A.~Strumia,
  %``Minimal flavour violation: An effective field theory approach,''
  Nucl.\ Phys.\  B {\bf 645}, 155 (2002)
  [arXiv:hep-ph/0207036].
  %%CITATION = NUPHA,B645,155;%%

%\cite{Cirigliano:2005ck}
\bibitem{Cirigliano:2005ck}
  V.~Cirigliano, B.~Grinstein, G.~Isidori and M.~B.~Wise,
  %``Minimal flavor violation in the lepton sector,''
  Nucl.\ Phys.\  B {\bf 728}, 121 (2005)
  [arXiv:hep-ph/0507001].
  %%CITATION = NUPHA,B728,121;%%

\bibitem{Allanach:2008ib}
  B.~C.~Allanach, J.~P.~Conlon and C.~G.~Lester,
  %``Measuring Smuon-Selectron Mass Splitting at the CERN LHC and
  %Patterns of Supersymmetry Breaking,''
  Phys.\ Rev.\  D {\bf 77}, 076006 (2008)
  [arXiv:0801.3666 [hep-ph]].
  %%CITATION = PHRVA,D77,076006;%%

\bibitem{ArkaniHamed:1996au}
  N.~Arkani-Hamed, H.~C.~Cheng, J.~L.~Feng and L.~J.~Hall,
  %``Probing Lepton Flavor Violation at Future Colliders,''
  Phys.\ Rev.\ Lett.\  {\bf 77}, 1937 (1996)
  [arXiv:hep-ph/9603431];
  %%CITATION = PRLTA,77,1937;%%
%\bibitem{ArkaniHamed:1997km}
%  N.~Arkani-Hamed, J.~L.~Feng, L.~J.~Hall and H.~C.~Cheng,
  %``CP violation from slepton oscillations at the LHC and NLC,''
  Nucl.\ Phys.\  B {\bf 505}, 3 (1997)
  [arXiv:hep-ph/9704205].
  %%CITATION = NUPHA,B505,3;%%

\bibitem{Engelhard:2009br}
  G.~Engelhard, J.~L.~Feng, I.~Galon, D.~Sanford and F.~Yu,
  %``SPICE: Simulation Package for Including Flavor in Collider Events,''
  arXiv:0904.1415 [hep-ph].
  %%CITATION = ARXIV:0904.1415;%%



\end{thebibliography}
\end{document}